\documentclass[useAMS]{gGAF2e}

\renewcommand*{\Pi}{\varPi}
\renewcommand*{\Omega}{\varOmega}

\begin{document}
\doi{10.1080/03091920xxxxxxxxx}
 \issn{1029-0419} \issnp{0309-1929} \jvol{00} \jnum{00} \jyear{} 

\markboth{W.-C.~M\"uller and S.K.~Malapaka}{Role of helicities for the dynamics of turbulent magnetic fields}

\title{{Role of helicities for the dynamics of turbulent magnetic fields}}

 \author{WOLF-CHRISTIAN M\"ULLER$\dagger$ and  SHIVA KUMAR MALAPAKA$^* \dagger \ddagger$ \thanks{$^*$ Corresponding author. Email: kumarshiva@gmail.com; previously at LJLL, UPMC, 4 place Jussieu 75005 Paris, France
\vspace{6pt} }\\
\vspace{6pt} $\dagger$Max-Planck Institut f\"{u}r Plasmaphysik, Boltzmannstr. 2,\\
 D-85748 Garching bei M\"unchen, Germany\\  
$\ddagger$Department of Mathematics, University of Leeds,
Leeds, LS2 9JT, UK\\}

\vspace{6pt} \received{This is an Author's Accepted Manuscript of an article published in  Geophysical \& Astrophysical Fluid Dynamics, First Published online :  01 Jun 2012. Print version:  Volume 107, Issue 1-2, 2013 Special Issue: From Mean-Field to Large-Scale Dynamos [copyright Taylor \& Francis], available online at: http://www.tandfonline.com/10.1080/03091929.2012.688292}

\maketitle

\begin{abstract} 
Investigations of the inverse cascade of magnetic helicity are conducted with
pseudospectral, three-dimensional
direct numerical simulations of forced and decaying incompressible magnetohydrodynamic turbulence.
The high-resolution simulations which allow for the necessary scale-separation show that 
the observed self-similar scaling behavior of magnetic helicity and related quantities
can only be understood by taking the full nonlinear interplay of velocity and magnetic 
fluctuations into account. With the help of the eddy-damped quasi-normal Markovian approximation
a probably universal relation between kinetic and magnetic helicities is derived that closely 
resembles the extended definition of the prominent dynamo pseudoscalar $\alpha$.
This unexpected similarity suggests an additional nonlinear quenching mechanism of the 
current-helicity contribution to $\alpha$. 


\end{abstract}


\section{Introduction} Understanding large-scale magnetic structure
 formation in the Universe is one of the challenging problems in
 modern astrophysics.  
In this context, mean-field dynamo theory is a prominent 
approach \citep{mo78,bis03,bransub05}. 
Based on a homogenization formalism, it describes the generation of large-scale magnetic fields
by smaller-scale turbulent fluctuations of a magnetofluid. As a result, this classical two-scale
closure \citep{krs80} yields, next to a turbulent diffusivity, a scalar, $\alpha\sim \tau H^K$, that expresses 
the nonlinear interaction of large-scale field and smaller-scale turbulence.   
Here, $\tau$ stands for a correlation time of the turbulent fluctuations and 
 $H^{K} =
 ({1}/{2V})\int_V{\bm {v\cdot\omega}}\,{\mathrm d}V$ is the kinetic helicity
of the associated velocity field ${\bm v}$ with $V$ being the volume under consideration and $\boldsymbol{\omega}=\nabla\times {\bm v}$ defining the vorticity.
Statistical closure theory \citep{pouq76} more specifically the eddy-damped quasi-normal Markovian (EDQNM) approximation, suggests a more
complex expression, $\alpha\sim\tau(H^K-H^J)$, that introduces the current helicity 
$H^J=({1}/{2V})\int_V{\bm {b\cdot j}}\,{\mathrm d}V$ with ${\bm j}$ denoting the 
electric current density, see also \citep{bl02,fi02,subbran04,bransub05}. 
Its name is actually misleading as $H^J$ expresses the helicity of the magnetic field
and is in this respect a close relative of the kinetic helicity and, furthermore, also 
proportional to the total resistive dissipation rate of magnetic helicity (see below).

While $H^K$ is ideally conserved and is spectrally cascading towards
smaller scales in the inertial range of three-dimensional
Navier-Stokes turbulence, the current helicity has apparently no
comparable significance for turbulent dynamics apart from its meaning
for the turbulent dynamo.  However, with electric current
${\bm j}={\bm \nabla}\times{\bm b}=-\Delta {\bm a}$, magnetic field
${\bm b}={\bm \nabla}\times{\bm a}$ 
and magnetic vector potential ${\bm a}$
(both dimensionless), a link to an ideal
invariant of three-dimensional incompressible magnetohydrodynamics (MHD)
emerges through the magnetic helicity, $H^{M} =
({1}/{2V})\int_V{\bm {a\cdot b}}\,{\mathrm d}V$. 
This quantity characterizing the topology of the magnetic field \citep{mo69}
is prone to an inverse cascade. The cascade is a robust nonlinear
mechanism that creates large scale order out of the chaotic randomness
of small-scale magnetic turbulence presupposing a sufficient
separation of large and turbulent small scales in the system in
combination with a small-scale supply of magnetic helicity.

The present work is motivated by the potential importance of magnetic
helicity for the dynamics of large-scale dynamo configurations. This
is not to be confused with the related issue of the effect of boundary
conditions on the magnetic helicity evolution and the consequences for
the dynamo process, a topic that has been subject of a number of
investigations \citep[see, e.g.,][and references therein]{bra09}.  In this work,
an idealized system, homogeneous incompressible MHD turbulence with triply
periodic boundary conditions, is investigated by three-dimensional
direct numerical simulations in combination with statistical closure
theory.

\section{Model equations and numerical Setup} 
The  dimensionless incompressible MHD equations giving a concise single-fluid description of a plasma are
\begin{subequations}
\label{solcon}
 \begin{equation}
 \partial_{t}{\boldsymbol{\omega}} =
 {\bm \nabla}\times({\boldsymbol{ v}}\times{\boldsymbol{\omega}} - {\boldsymbol{ b}}\times{\boldsymbol{j}})+
 {\mu}_{{n}}(-1)^{n/2-1}\nabla^{{n}}
 \boldsymbol{\omega}+{\boldsymbol{ {F_{v}}}}+{\lambda}\Delta^{-{1}}\boldsymbol{\omega}\,,
 \end{equation}
 \begin{equation}
\partial_{t}{\bf{b}} = {\bm \nabla}\times({\boldsymbol{v}} \times {\boldsymbol{b}}) +
 {\eta}_{{n}}(-1)^{n/2-1}\nabla^{{n}}
 {\boldsymbol{b}}+{\boldsymbol{{F_{b}}}}+{\lambda}\Delta^{-{1}}{\boldsymbol{b}}\,,
\end{equation}
\begin{equation}
 {\boldsymbol{ {\nabla\cdot v}}} = {\boldsymbol{ {\nabla\cdot b}}}= {0}.
\end{equation}
\end{subequations}
 Relativistic effects are neglected and the mass density is
 assumed to be 
unity throughout the system. Other effects
 such as convection, radiation and rotation are also neglected.
 Direct numerical simulations are performed by solving the 
 set of model equations by a standard pseudospectral method \citep{can88}
 in combination with leap-frog integration 
 on a cubic box of linear size $2\pi$ that is discretized with $1024$ collocation points 
 in each spatial dimension. 
 Spherical mode truncation is used for alleviating aliasing errors. 
 By solving the equations in Fourier space, the solenoidality of ${\bm v}$ and ${\bm b}$
 is maintained algebraically.

To observe clear signatures of an inverse cascade of magnetic helicity 
the system has to contain a source of this quantity at small scales.
This is achieved in two different ways resulting in two main configurations:
a driven system and a decaying one. In the driven case, the forcing terms 
${\bm {F_v}}$ and ${\bm {F_b}}$ are delta-correlated random processes 
acting in a band of wavenumbers $203\leq k_0\leq 209$. They create a 
small-scale background of fluctuations with adjustable amount of magnetic and 
kinetic helicity. The results reported in this paper do not change if kinetic helicity injection
is finite. 
The theoretical results presented in the following do not depend on 
the  setup of the forcing as they presuppose an existing self-similar distribution of
energies and helicities. For obtaining such spectra in numerical experiments 
the magnetic source term ${\bm {F_b}}$ is necessary while a finite momentum source ${\bm {F_v}}$ 
speeds up
the spectral development significantly.
In the decaying case
the forcing terms are set to zero and the initial condition represents an
ensemble of  smooth and random fluctuations of maximum magnetic helicity 
with respect to the energy
content (see below) and a characteristic wavenumber $k_0=70$.

To reduce finite-size effects,
the simulations are run for $6.7$ (forced) and $9.2$ (decaying) large-eddy
turnover times of the system, respectively. The time unit is defined using the system size
and its total energy. Additionally, a large scale energy sink 
$\lambda\Delta^{-1}$ with $\lambda=0.5$ is present 
for both fields. In the decaying case $\lambda=0$. 
The hyperdiffusivities ${\mu}_{{n}}$ and ${\eta}_{{n}}$ are 
dimensionless dissipation coefficients of order $n$ (always even in these simulations), with $n=8$ in both runs. 
They act like higher-order
realizations of viscosity and magnetic diffusivity, respectively. 
The magnetic hyperdiffusive Prandtl number
${Pr_m}_n={\mu_n}/{\eta_n}$ is set to unity.

The initial conditions to these simulations are smooth fluctuations with 
random phases having a Gaussian
energy distribution  peaked around $k_0$ in the decaying and the forced cases.
Magnetic and kinetic helicity of the initial state can be controlled 
in the same way as for the forcing terms \citep[cf.][]{bismul00}.
The initial/force-supplied ratio of kinetic to magnetic energy is unity
with an amplitude of 0.05 in the forced
 case and an amplitude of unity in the decaying case.  Hyperviscosity
 of order $n=8$ is chosen in the simulations to obtain sufficient scale-separation. 
It is difficult to define an unambiguous Reynolds number owing to the use
 of hyperviscosity \citep[][and the references therein]{mal09}. With
 the above mentioned simulation set up, the equations are solved both
 for decaying and forced cases separately and the results obtained are
 discussed below.   

 \section{Simulation results} 
Using the simulation setup described in the previous section, 
inverse cascading of magnetic helicity with a clear
 scale separation between large and small scales is established
 in both forced and decaying cases for wavenumbers $k<k_0$.
This is indicated by the spectral flux, $\Pi^{H^{M}}_k=
 \int^k_0{\mathrm d}k' \int {\mathrm d}\Omega \bigl[{\tilde{\bm b}}^*{\bm \cdot}\,(\widetilde{{\bm v}\times{\bm b}})\bigr]_{|{\bm k}'|=k'}$,
 in both cases depicted in figure 1(b) and 
taken at $t=6.7$ and $9.2$, respectively as dissipation of magnetic helicity is negligible (see figure 1(a)). The tilde indicates Fourier transformation and $^*$ stands for complex 
conjugate.
The inverse flux in the driven case is constant over a significant spectral interval, indicating equilibrium of 
source and sink, while the temporal decay of the magnetic helicity reservoir in the decaying case
is reflected by the associated non constant inverse flux. In both cases the characteristic 
wavenumber of the $H^M$-source can be identified as the separation between inverse and direct 
flux regions.
The spectral flux of magnetic helicity has been extensively studied in earlier 
numerical simulations \citep[see, e.g.,][]{brand01,alexi06}. These works,
however, are lacking the necessary scale separation to observe self-similar 
scaling laws. 
\begin{figure}
a) \includegraphics[width=0.5\textwidth]{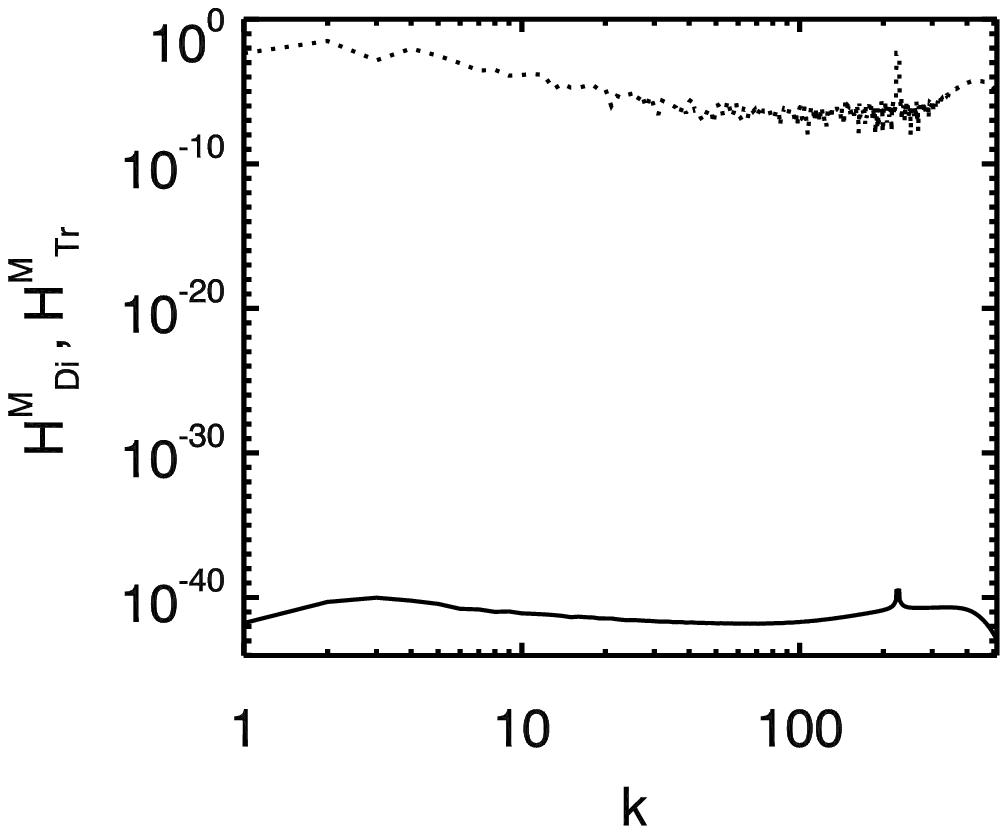} b)\includegraphics[width=0.5\textwidth]{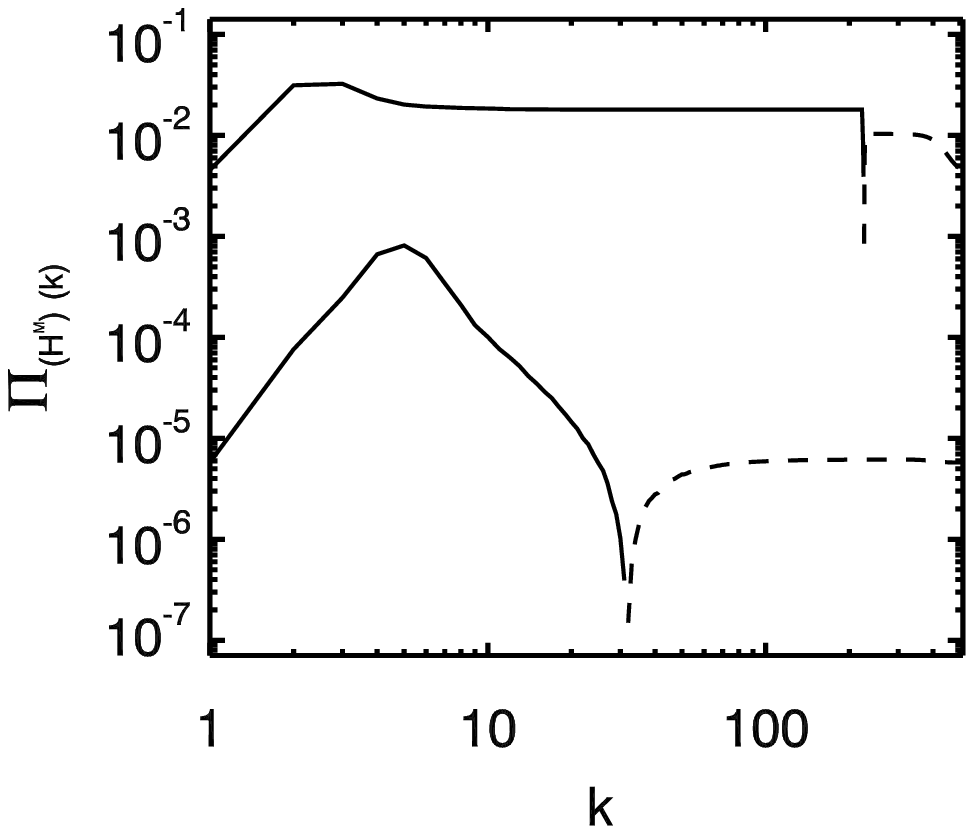}
\caption{ (a) Transmission $H^{M}_{Tr}$ = ${\tilde{\bm b}}^{*}{\bm \cdot}\,(\widetilde{\overline{{\bm v}\times{\bm b}}})$ (dotted line) and dissipation $H^{M}_{Di}$ = ${{\eta_n}{{k}^{6}}}\tilde{\bm b}^{*}{\bm \cdot}\,\tilde{\bm j}$ (solid line) of magnetic helicity space-angle-integrated in Fourier space in the forced case (similar for the decaying case, not shown). (b) Spectral flux of magnetic helicity in the forced (top) and decaying cases (bottom), dashed curves: direct flux, solid curves: inverse flux.}
\label{fig:maghelflux}
\end{figure}
The spectrum of magnetic helicity exhibits 
scaling behavior $\sim k^q$ with $q\approx -3.3$ and 
 $q\approx {-3.6}$ (forced and decaying case, respectively) which cannot be explained by 
the straightforward constant-flux reasoning \`a la Kolmogorov adopted in \cite{pouq76} to interpret their EDQNM results. 

In fact, the involved dimensional argument (Alfv\'enic units), $[H^M_k]=L^4/T^2$ (spectrum), $[\varepsilon_M]=L^3/T^3$ (spectral flux), 
in combination with the assumption of spectral self-similarity, $H^M_k\sim \varepsilon_M^a k^b$, yields $a=2/3$, $b=-2$, but 
does not explicitly include the nonlinear interaction of velocity and magnetic fields \citep[see also][]{bis03}. Here `$L$' represents length and `$T$' the time. As a first step in the 
necessary refinement of the theoretical modeling additional consideration
of the kinetic helicity $H^K_k$ seems appropriate.

As a consequence of the inverse spectral transfer of magnetic helicity, 
all magnetic quantities should inherit the 
observed spectral inverse transfer property. 
This is indeed the case for the magnetic energy, the electric current density,  
and the current helicity. These quantities also show self-similar scaling
that however, differs to some degree between the two investigated configurations. 
It is particularly interesting, that the residual helicity 
$H^R $ = $\left|H^V \ - \ k^2 \,H^M\right|$, also shows self-similar scaling with
$q\approx {-1.4}$ and $q\approx {-1.8}$ in the forced and decaying cases respectively \citep[see][for further details]{mal09}.
The interaction of the magnetic field with the velocity in a progressing inverse cascade of 
magnetic helicity appears to be of importance for a better understanding of the observed 
scaling laws. At high Reynolds numbers, the process of large-scale magnetic structure formation
by the inverse cascade is accompanied by a continuous stirring of the velocity field caused 
by the expanding magnetic field structure. 
The magnetic stirring of the MHD-fluid leads to a transfer of magnetic to
kinetic energy and generates ever larger velocity fluctuations. These also show
self-similar scaling, as, for example, reflected by the kinetic helicity spectrum
with $q\approx -0.4$ (forced case) and $q\approx 0.4$ (decaying case).

With regard to the finding \citep[e.g.][]{alexi06} of the pronounced 
spectral non-locality of the nonlinear interactions underlying 
$\Pi_k^{H^M}$ a few words about the physical picture of the inverse 
cascade are in order.   
The cascading process is realized as 
a merging of positively-aligned and thus mutually attracting current carrying structures
\citep[cf.][]{bisbrem93}.
It is not necessary that the structures grow in size as they indeed do in the decaying case,
as long as the corresponding current densities increase. This is observed in the simulation with 
small-scale forcing.
As there is no obvious fluid-dynamical constraint on the merging of two current filaments with
regard to their size, this picture is consistent with a spectrally non-local 
inverse cascade of magnetic helicity. 

\section{Spectral relationship between kinetic and magnetic helicities} 
A link between kinetic and magnetic helicities can be constructed with the help of dimensional analysis of the magnetic helicity evolution equation in the EDQNM approximation, a statistical closure model discussed \citep[e.g., in][]{pouq76}. Such an approach
was successful earlier, in describing the turbulent residual energy spectrum, $E^R_k=|E_k^M-E_k^V|$ yielding $E^R_k\sim kE_k^2$ \citep{muegra05} with $E_k=E^M_k+E_k^K$, which also turns out to be valid in the present simulations, where $E_k^M$ and $E_k^V$ are magnetic and kinetic energies respectively. 

Assuming that the most important nonlinearities involve the turbulent velocity and stationarity 
of the spectral scaling range of $H_k^M$, a dynamical equilibrium of turbulent advection and 
the $H^M$-increasing effect
of helical fluctuations is proposed.  This can be formulated straightforwardly 
using the corresponding dimensionally approximated nonlinear terms from the EDQNM model
\citep[for a more detailed derivation see][]{mulmal12},  yielding
\begin{equation}
H^K_k\sim \bigl({E^K_k}/{E^M_k}\bigr) k^2 H_k^M\,. 
\label{mainres}
\end{equation} 
This statement about the spectral dynamics of kinetic and magnetic helicities 
(or, equivalently, kinetic and current helicities since $H^J_k\sim k^2H^M_k$) 
is also valid for $E_k^K/E_k^M\neq 1$.
The agreement of relation (\ref{mainres}) with the numerical experiments is however
significantly improved by a modification (relation (\ref{main2res}) below) whose justification
is beyond the scope of the presented equilibrium ansatz which basically assumes spectral 
locality of the inverse cascade
\begin{equation} 
H^K_k\sim \bigl({E^K_k}/{E^M_k}\bigr)^2H_k^J\,.
\label{main2res}
\end{equation}
Relation (\ref{main2res}) is a significant improvement over the earlier relations of similar kind \citep{pouq76,pouq10,mulmal10}. This is shown in figures 2(a) and 3(a), where $\Theta=(E^K_k/E^M_k)^\gamma H^J_k/H^K_k$ is shown with $\gamma = 0,1$ and $2$ (corresponding to $\Theta$, $\Theta _1$ and $\Theta _2$) for the forced and decaying cases respectively. It is remarkable that relation (\ref{main2res}) is only fulfilled in wavenumber intervals where the flux of magnetic helicity is spectrally constant. 

\begin{figure}[h]
 \includegraphics[width=0.95\textwidth]{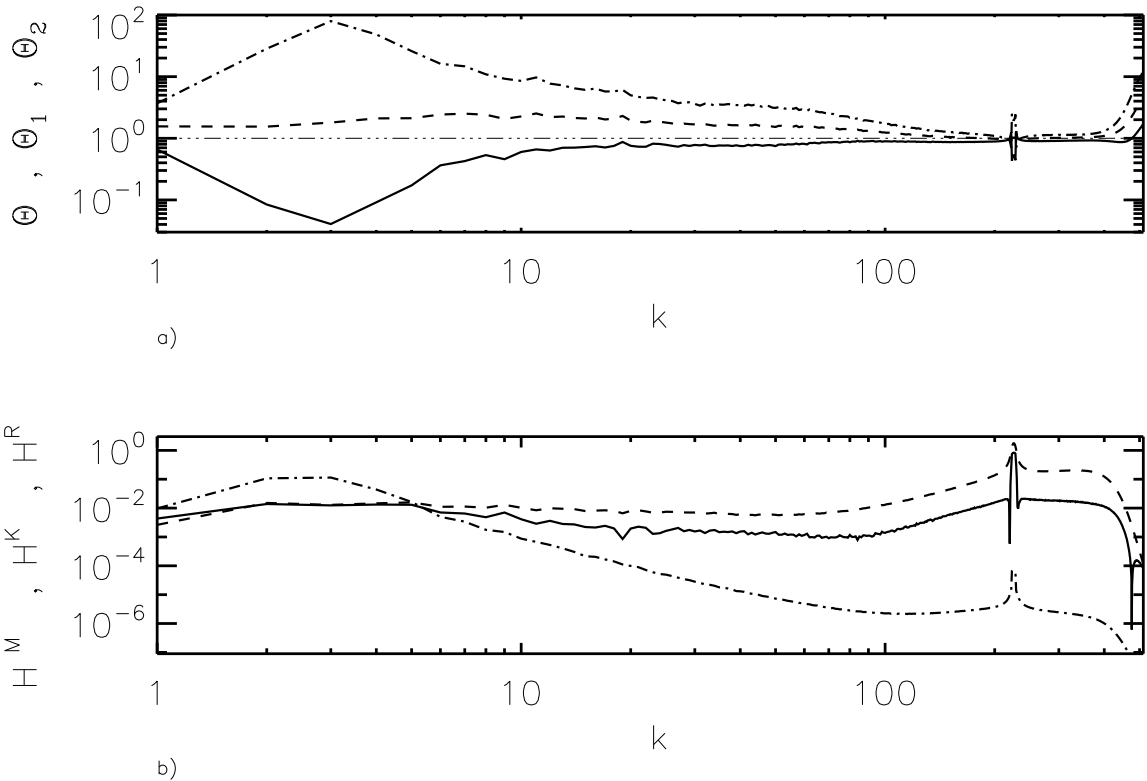}
\caption{Plots of relation (\ref{main2res}), and kinetic, magnetic and residual helicities for forced turbulence at t=6.7.
(a) Relation (\ref{main2res}) $\Theta=(E^K_k/E^M_k)^\gamma H^J_k/H^K_k$. $\gamma=0$ (dash-dot curve) $\gamma=1$ (dashed curve), and $\gamma=2$ (solid curve). (b) Magnetic helicity (dash-dot curve), kinetic helicity (dashed curve) and residual helicity (solid curve).}
\end{figure}
\begin{figure}[h]
 \includegraphics[width=0.95\textwidth]{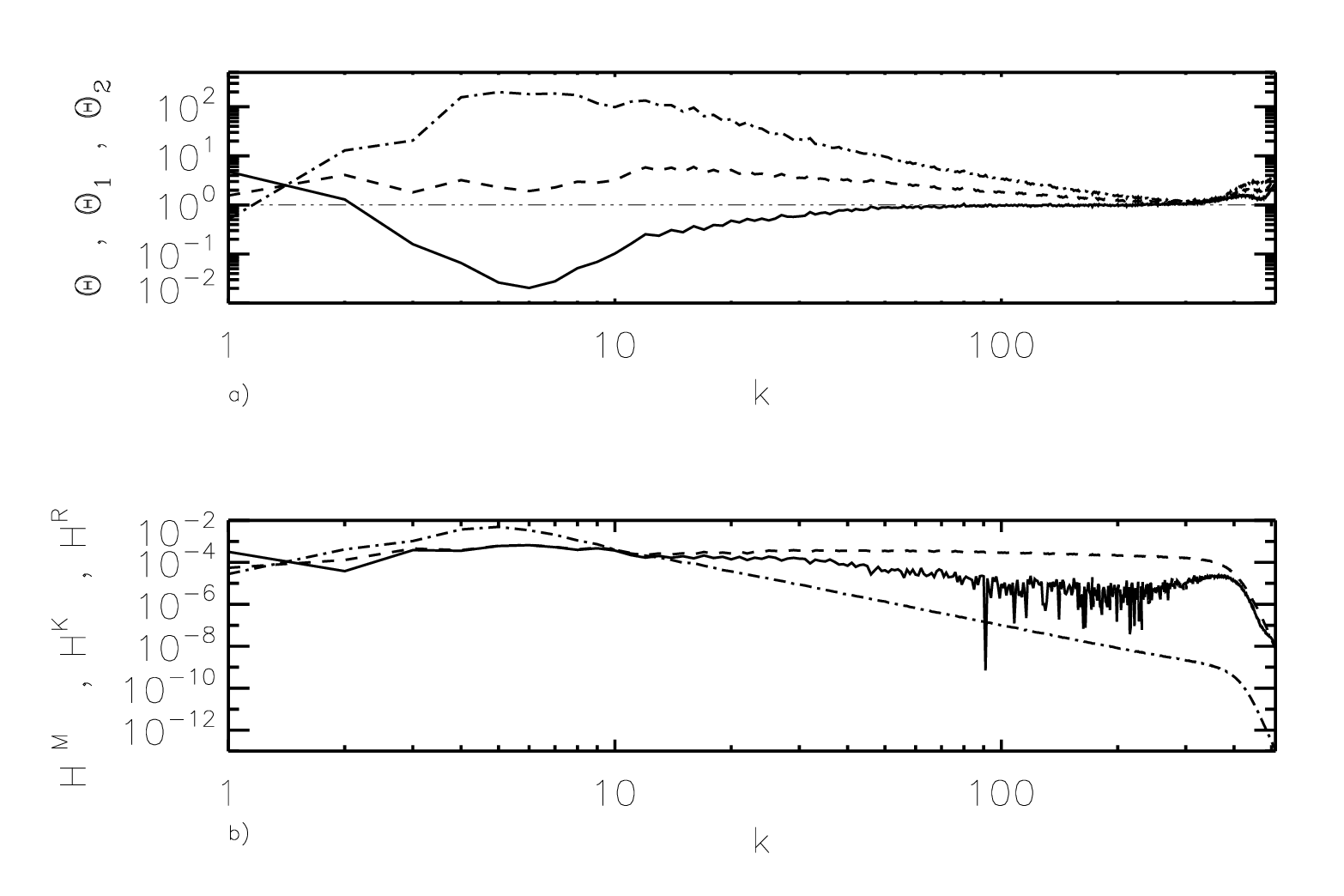}
\caption{Plots of relation (\ref{main2res}), and kinetic, magnetic and residual helicities for decaying turbulence at t=9.2.
(a) Relation (\ref{main2res}) $\Theta=(E^K_k/E^M_k)^\gamma H^J_k/H^K_k$. $\gamma=0$ (dash-dot curve) $\gamma=1$ (dashed curve), and $\gamma=2$ (solid curve). (b) Magnetic helicity (dash-dot curve), kinetic helicity (dashed curve) and residual helicity (solid curve).}
\end{figure}
 This relation brings back the ratio of energies (kinetic to magnetic)
 into the picture, which, under the assumption of equipartition of
 energies was ignored in previous work \citep{pouq76}, while
 linking the magnetic/current and kinetic helicities. 
 Another interpretation for this expression is the
 partial Alfv\'enization of the turbulent flow
 \citep{pouq10}. Further, it also highlights the influence of kinetic
 helicity in the inverse cascade of magnetic helicity.

The relation (\ref{main2res}) belongs to a class of probably highly universal expressions
which are based statistically on the quasi-normal approximation of nonlinear fluxes. 
It is interesting to note that relation (\ref{main2res}) also allows to determine the spectral scaling exponent of 
magnetic helicity  from astronomical current helicity measurements using vector magnetograms 
\cite[see, e.g.,][and references therein]{bransub05} if kinetic and magnetic energy spectra are also measurable or can be estimated with sufficient 
accuracy.

The modification of the current helicity contribution present in relation (\ref{main2res}) suggests a 
corresponding modification to the residual helicity, $H^R= H^K-H^J$, and accordingly 
to the mean-field dynamo $\alpha$. This, however, has to be taken with care as 
the present simulations are energetically dominated by the magnetic field although 
the modifying factor $(E^K/E^M)^2$ should compensate for this. 
Figures 2(b) and 3(b) allow us to roughly estimate the respective scale-dependent 
influence of 
kinetic and magnetic helicity on the modified residual helicity. { The spectrum of residual helicity closely follows the spectral kinetic helicity with growing systematic deviations due to the 
influence of magnetic helicity at large wavenumbers, in both cases. 
Thus, the modified residual helicity complies with the earlier definitions of $\alpha$ \citep{krs80,pouq76}} at large scales.
\section{Conclusions} 
In high-resolution direct numerical simulations of forced and decaying magnetically 
helical homogeneous MHD turbulence, the nonlinear dynamics of 
active inverse cascade of magnetic helicity is studied. 
The simulation results, in particular the 
observed self-similar spectral scaling of magnetic helicity which contradicts 
an earlier theoretical explanation \citep{pouq76}, motivate the consideration 
of velocity field characteristics for the nonlinear evolution of 
this purely magnetic quantity. This is done with the help of statistical closure
theory yielding a possibly universal relation between kinetic and current helicities.
The relation is corroborated by the numerical results.
Its form, $H_k^K-(E^K_k/E^M_k)^2H^J_k\sim \text{constant}$, closely resembles the extended definition
of the pseudo-scalar $\alpha\sim H^K_k-H^J_k$ known from mean-field dynamo theory.
The inverse cascade of magnetic helicity is neither a dynamo in itself (as dimensionally 
$H^M_k\sim k^{-1}E^M_k$) nor even a turbulent cascade in the strict sense, but is a spectral transport process \citep{mulmal12}. It can as a robust and efficient spectral transporter, nevertheless, 
play a role in the actual realization of turbulent large-scale dynamos like the $\alpha$-dynamo.
In this respect it is interesting that the newly obtained relation includes the squared
ratio of kinetic and magnetic energies. This leads to a purely 
nonlinear quenching of the current helicity contribution to $\alpha$ that has no direct 
connection to the dynamo-quenching mechanisms considered so far (of order $(E^M)^{-1}$) in the literature
which are seemingly consequences of a combination of boundary conditions and the 
approximate conservation of magnetic helicity.
In this context, it is encouraging that \cite{rhebran10}
for a homogeneous mean flow with Roberts forcing using a test field method
observe $\alpha$-quenching with an  $(E^M)^{-2}$ signature. The comparison with
this work assumes equivalence of 
their imposed mean field with the root-mean-square large-scale
magnetic fluctuations in the present simulations.

The present relation (\ref{main2res}) needs further investigation 
as it is an additional possible mechanism for dynamo quenching. 
This new link between kinetic and magnetic helicity in the inverse cascade of magnetic 
helicity has to be verified in more complex numerical setups such as mean field dynamos, as
well as 
anisotropic 3D-MHD and isotropic 3D-MHD turbulence with different initial conditions and 
forcing mechanisms. \vspace{2cm} \\
{\bf{Acknowledgments}}\\
SKM thanks B. Despres of JLLL, UPMC, Paris, VI, and CNRS
for the financial support they provided to attend the R\"adler
fest, A.~Brandenburg and his NORDITA team for
hosting him at this fest, as well as
A.~Busse currently in Southampton, UK and D.~Hughes at
University of Leeds, UK for their help and useful discussions.
The authors want to thank U.~Frisch for useful remarks.



\label{lastpage}
\end{document}